\documentclass[]{eptcs} 
\providecommand{\event}{Workshop on Hybrid Autonomous Systems 2013} 

\usepackage[centertags]{amsmath}   
\usepackage{amssymb}               
\usepackage{latexsym}              
\usepackage{mathrsfs}

\usepackage{graphicx}
\usepackage{epsfig}
\usepackage{subfigure}
\usepackage{epic}
\usepackage{eepic}
\usepackage{color}

\usepackage[T1]{fontenc}

\DeclareMathOperator{\sign}{sign}
\DeclareMathOperator{\real}{Re}
\DeclareMathOperator{\imag}{Im}

\usepackage[matrix,arrow,cmtip,rotate,curve]{xy}
\SilentMatrices                

\input xy
\xyoption{all}

\newdir{c}{{}*!/-5pt/@^{(}}    

\newdir{d}{{}*!/-5pt/@_{(}}

\title{A Simple Stochastic Differential Equation with Discontinuous Drift}
\author{Maria Simonsen, John Leth, Henrik
  Schioler
\institute{Dept. of Electronic Systems, Automation
    and Control \\ University of Aalborg, Denmark}
\email{\{msi,jjl,henrik\}@es.aau.dk}
\and
Horia Cornean 
\institute{Dept. of Mathematical Sciences \\ University of Aalborg, Denmark}
\email{cornean@math.aau.dk}
}
\def\titlerunning{A Simple Stochastic Differential
  Equation with Discontinuous Drift}
\def\authorrunning{M. Simonsen, J. Leth, H. Schioler \& H. Cornean}

\begin{document}
\maketitle

\begin{abstract}
In this paper we study solutions to stochastic differential equations (SDEs) with discontinuous drift. 
We apply two approaches: The Euler-Maruyama method and the Fokker-Planck equation and show that a candidate density function based on the Euler-Maruyama method approximates a candidate density function based on the stationary Fokker-Planck equation.  Furthermore, we introduce a smooth function which approximates the discontinuous drift and apply the Euler-Maruyama method and the Fokker-Planck equation with this input. 
The point of departure for this work is a particular SDE with discontinuous drift. 
\end{abstract}


\section{Introduction}

Since the pioneering work by Einstein on Brownian motion \cite{ein05},
stochastic differential equations (SDEs) have been intensively
studied, with the foundation for SDEs developed by Ito and
Stratonovich, e.g.\ see \cite{ito51,str66} and references
therein. Well-developed theories for various sub-disciplines of SDEs
have been developed ever since, such as stability-  and
control theory \cite{kha12,kus67,kus71}. Lately, SDEs have been used in the
generalization of hybrid dynamical systems \cite{buj06,JohnLygeros,mao06}.
Moreover, today a variety of applications and numerical methods exist
for SDEs, see \cite{oks03,klo92} and references therein.

Comment to the above theories is that they are all developed under
(weak) regularity conditions on the drift and diffusion coefficients.
Such conditions are, of course, necessary in order to develop an
applicable/operational theory. However, it is also of interest to
study special cases (or classes) where the standard regularity
conditions fails but fundamental properties of the SDEs are still
valid such as existence and uniqueness of solutions. As an example, we
recall that a necessary condition for existence of solutions for an
(deterministic) ordinary differential equation (ODE), $\dot{x}=f(x)$,
is that the right-hand side of the ODE, that is the vector field $f$,
should be continuous in the state variable $x$. However, the Cauchy
problem $\dot{x}=\sign(x),~x(0)=x_0$, with $\sign$ denoting the $\sign$-function, has a solution for all initial values $x_0$, despite the
fact that $\sign$ is discontinuous at zero (we remark that $\sign(0)=0$ by
definition, if this was not the case no solution would exist at
$x_0=0$). As discontinuous functions often appear in applications, ODEs
such as $\dot{x}=\sign(x)$ should and have been studied, see \cite{fil88}
and references therein.

In this work in progress we initiate a study of SDEs with
discontinuous drift. This should be seen as a part of a larger scope
with focuses on the intersection between SDEs and switching dynamics,
a field only scarcely explored so far, see \cite{bri08} and references
therein for related work. Within this larger scope theoretical questions to
be answered are how to define solutions to SDEs with state dependent
switching (in particular with discontinuous drift), when solutions
cannot be patched together of segments of positive time duration and
can any applicable results be obtained by applying the Euler Maruyama 
method \cite{may,DesHig}, which is a simple time discrete approximations technique, to SDEs whose drift do not meet the regularity conditions? Here,
we exemplify these problems by studying solution candidates to a
particular SDE having discontinuous drift coefficient. More precisely,
we consider the SDE
\begin{equation}
dx_t = -k\sign(x_t)dt + dB_t,
\label{eq:disSDE}
\end{equation}
with $k$ a control gain, and ask how a solution can be defined. 

The systems treated in \cite{buj06} are assumed to have non-zeno execution in finite time. A local behavior of the Wiener process $B_t$ is that in every time interval of the form $[0,\epsilon)$, with $\epsilon >0$, $B_t$ has infinitely many zeros, that is, the process which fulfills 
\begin{eqnarray*}
 dx_t = dB_t
\end{eqnarray*} 
crosses zero infinitely many times, see \cite{Arnold}. In \eqref{eq:disSDE} the process $x_t$ is forced to proceed against zero, so we conjecture that the solution to \eqref{eq:disSDE} (if it exists) also crosses zero infinitely many times in finite time. 
 
We use several methods in the attempt to give a meaningful/operational
definition of solutions to \eqref{eq:disSDE} based on density functions and their probabilistic properties. 
We start by using the Euler-Maruyama method to approximate numerical solutions to \eqref{eq:disSDE}, which values are presented in histograms for particular time instant. These histograms can be considered as an approximation to the density function for solutions to \eqref{eq:disSDE}. We investigate the influence of the step-size and the control gain in the   
simulations.  
Furthermore, from the Euler-Maruyama method we obtain recursively defined density
functions which, under stationary conditions, appears to 
converge to the outcome
of the second approach which departure from the Fokker-Planck
equation. 
It is interesting to note that the candidate density function
obtained from the Fokker-Planck equation is derived under the
assumption of stationarity while the recursive approach has no such
assumptions. 
More precise, we obtain formulas which strongly indicate that
if such a stationary density function exists then it solves the
stationary Fokker-Planck equation.  

As a third method, we introduce an approximation to the $\sign$-function and apply both the Euler-Maruyama method and the Fokker-Planck equation to this. A comparison with the stationary density function which solves the Fokker-Planck equation is made.

Finally, we briefly mention one approach which relate to the
Euler-Maruyama method. Even though this intuitively should produce some
information, we have so far not been able to obtain any meaningful
results based on this method. It is included since it is believed
that it does in fact carry important information.

It is important to emphasize that the presented material is work in progress and that the heuristic presented here is an initial attempt to define meaningful candidate density functions to solutions to a particular SDE with discontinuous drift. It is clear that for future work the presented material have to be set in a formal mathematical frame including proofs which validate the various procedure used to obtain candidate density functions.


\subsection{A Stochastic Differential Equation} \label{SDE}

A general one dimensional SDE is given by
\begin{equation}
dx_t = b(t,x_t) dt + \sigma(t,x_t)dB_t,\quad x_0=c,
\label{eq:generaldiff}
\end{equation}  
where $x=x_t$ is an $\mathbb{R}$-valued stochastic process
$:[0,T]\rightarrow \mathbb{R}$, $b,\sigma:[0,T]\times \mathbb{R}
\rightarrow \mathbb{R}$ are the drift and diffusion coefficient of
$x$, $B=B_t$ is an $\mathbb{R}$-valued Wiener process, and $c$ is a random
variable independent of $B_t-B_0$ for $t\geq 0$. On $[0,T]$, existence
and uniqueness of a solution $x_t$, continuous with probability 1, to
\eqref{eq:generaldiff} is guaranteed whenever the drift $b$ and
diffusion $\sigma$ are measurable functions satisfying a Lipschitz condition
together with a growth bound, both uniformly in $t$ \cite[Theorem 5.2.1]{oks03}.

This paper focuses on the special case for \eqref{eq:generaldiff},
where $\sigma(t,x_t)=1$ and $b(t,x_t)=-k\sign(x_t)$ with $k>0$ a control
 gain and the $\sign$-function defined by
\begin{equation}
\sign(x) = \left\{ \begin{array}{ccc}
-1 & \mbox{ if } & x<0 \\ 0 & \mbox{ if } & x=0 \\ 1 & \mbox{ if } & x>0  
\end{array} \right. \;.
\label{eq:defsignfunction}
\end{equation}
Thus, we consider the SDE 
\begin{equation}
dx_t = -k \sign(x_t)dt + dB,\quad x_0=c. 
\label{eq:basisdiff}
\end{equation}
with $c$ given.

In the next section, the Euler-Maruyama method is applied to approximate solutions to \eqref{eq:basisdiff} and to investigate a theoretical methods to obtain candidate density functions for solutions to \eqref{eq:basisdiff}.

\section{The Euler-Maruyama Method} \label{EulerMaruyama} 
The Euler-Maruyama method is a simple time discrete approximation technique which is used to approximate solutions to SDEs of
the type given in \eqref{eq:generaldiff}, by discretizing the time
interval $[0,T]$ in steps $0 < t_1 < \cdots < t_n < t_{n+1} \cdots <
t_N$ with $N=\left\lceil \frac{T}{h} \right \rceil$, where $h =
t_{n+1}-t_n$ is the step-length. Each recursive step is determined via
the following method,
\begin{equation}
x_{n+1} = x_n + hb(t_n,x_n)+\sigma(t_n,x_n)W_n \;,
\label{eq:deterministiskStep32}
\end{equation}
where $x_{t_i}=x_i$ and $W_n=B_{t_{n+1}}-B_{t_{n}}$ is 
i.i.d. normal with mean zero and variance $h$, which we denote by $W_n \sim N(0,h)$.

Given an initial condition $x_0=c$, it is possible from
\eqref{eq:deterministiskStep32} to approximate a solution to
\eqref{eq:generaldiff} by determination of $x_1,x_2,\ldots,x_N$.  If
the drift and diffusion coefficient in \eqref{eq:generaldiff} are 
measurable, satisfy
a Lipschitz condition and a growth bound, the Euler Maruyama method
guarantee strong convergence to the solution of
\eqref{eq:generaldiff}, \cite[Theorem 9.6.2]{klo92}. Hence for
SDEs with discontinuous drift we can, in general, not expect the Euler
Maruyama method to produce meaningful results. Nevertheless, we will
in the sequel apply this method to the special case
\eqref{eq:basisdiff} in order to obtain candidate
solutions.

\subsection{Analysis of the Deterministic Step} \label{sec:deterministiskStep}

Application of the Euler-Maruyama method to the SDE in
\eqref{eq:basisdiff} gives the recursive step
\begin{equation}
x_{n+1} = x_n - h k \sign(x_n)+W_n \;.
\label{eq:deterministicStep}
\end{equation}
Now, if $x_n> 0$, we have
\begin{equation*}
x_{n+1} = x_n -hk + W_n.
\label{eq:positiveside}
\end{equation*} 
Since $W_n \sim N(0,h)$, the expectation is that $x_{n+1} \in
[x_n-h(k+1),x_n-h(k-1) ]$ in most of the simulations. From this, we
expect after a finite time $0<t<\infty$ that there exists $N\in
\mathbb{N}$ such that $x_{n+N}\leq 0$. Similar result is obtained if
$x_m<0$, then we expect that there exists  $M\in \mathbb{N}$
such that $x_{m+M}\geq 0$. The influence from the control gain $k$
determines how quick the evolution of the sequence $\{x_n\}_{n\geq 0}$ switches around zero.  In other words, a big $k$ minimizes the
influence of the random variable $W_n$.

The Euler-Maruyama method is easy to implement in software, so following we have applied Matlab to simulate solutions to \eqref{eq:basisdiff}. 

\subsection{Numerical solutions to a SDE with Discontinuous Drift} \label{simulation} 
We consider the recursive step in \eqref{eq:deterministicStep} and simulate the evolution of the stochastic process. 
For all simulations the initial condition is chosen to be $x_0 = 0$ and the
considered time interval is $[0,T]$ where $T=1$. The step-length is
$h$ such that the number of simulated steps is $N=\left\lceil
\frac{T}{h} \right\rceil$. All simulations are repeated $500$ times and histograms of the resulting values of $x_T$ are presented. 

\begin{figure}%
\centering
\parbox{2.6in}{%
\includegraphics[trim=5cm 10cm 5cm 10cm,
width=0.40\textwidth]{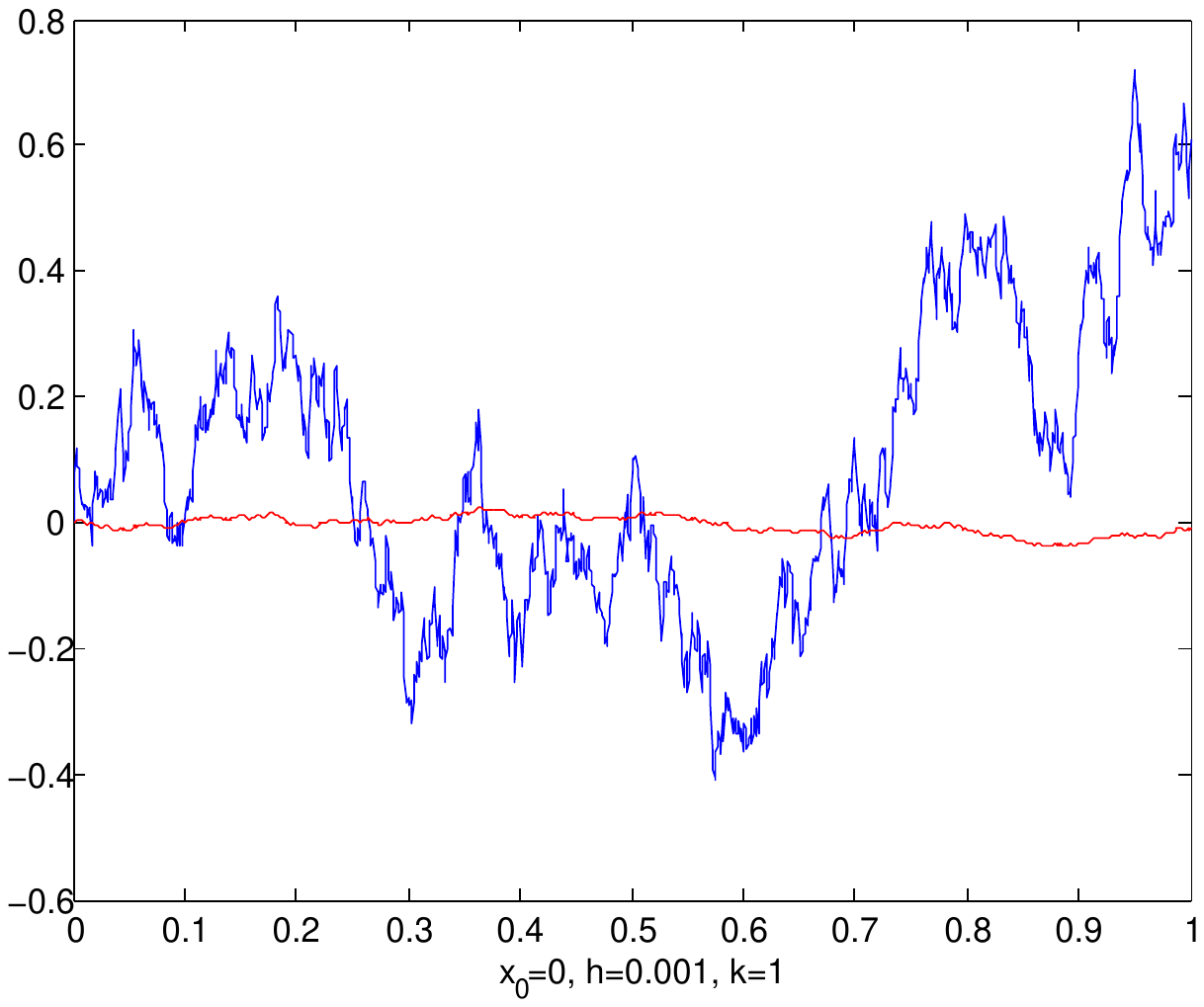}
\caption{The blue graph shows one realization
  out of $500$ simulations while the red graph represents the average
  of $x_t$ for $0\leq t \leq T$.}
\label{fig:EnkelSimuIni0}	}
\qquad 
\parbox{2.6in}{
\includegraphics[trim=5cm 10cm 5cm 10cm,
width=0.40\textwidth]{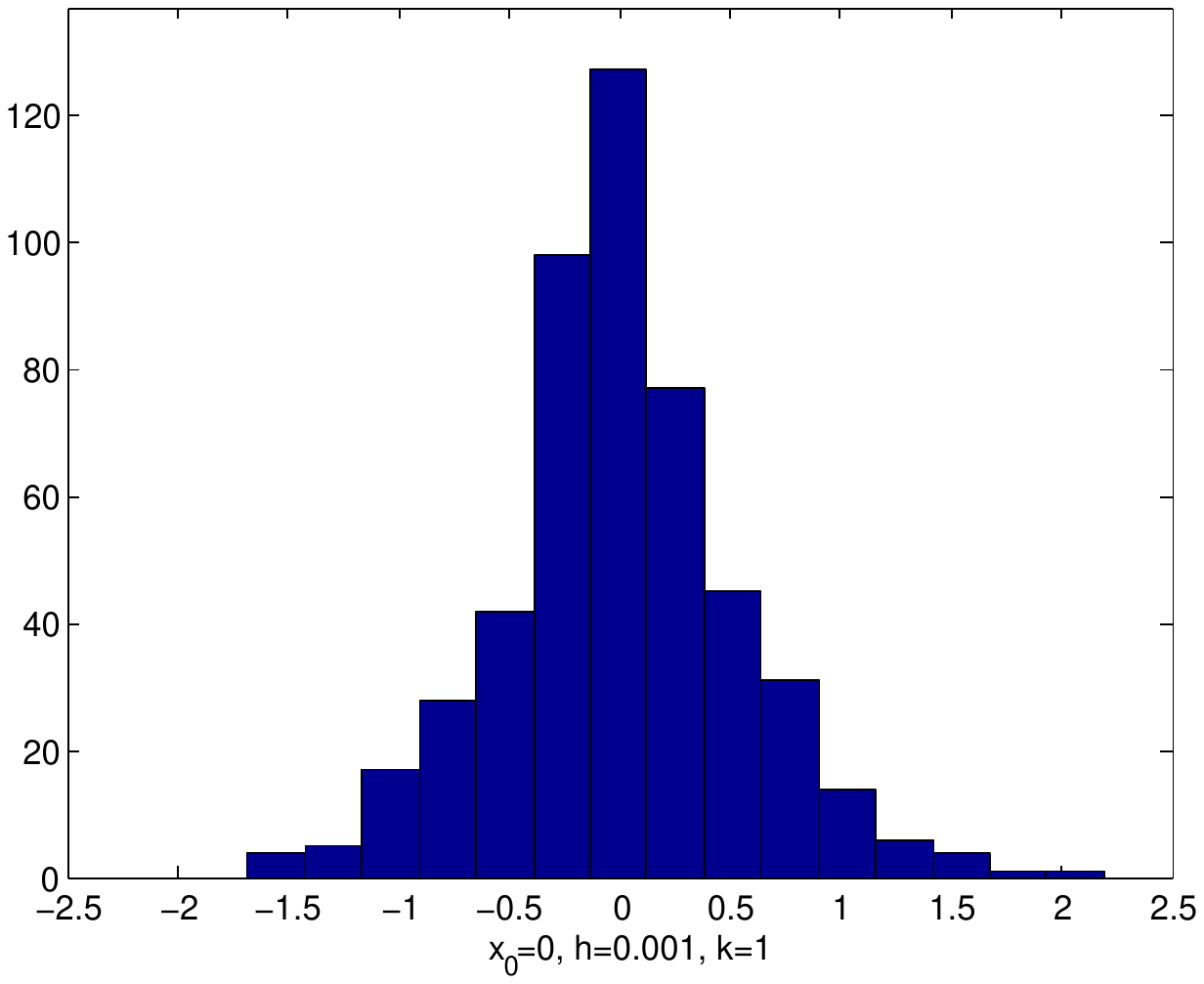}
\caption{Resulting histogram of $500$ simulations of $x_T$.}
\label{fig:HistoIni0}}
\end{figure}

In figure \ref{fig:EnkelSimuIni0}, one realization of a solution to the SDE in \eqref{eq:basisdiff} is shown
together with the average values of all the $500$ simulations in the time interval $[0,T]$. The
average of $x_t$ is close to zero for all $0\leq t \leq T$.

Figure \ref{fig:HistoIni0} shows the resulting histogram of $x_T$
including $500$ simulations.

\begin{figure}%
\centering
\parbox{2.6in}{%
\includegraphics[trim=5cm 10cm 5cm 10cm,
width=0.40\textwidth]{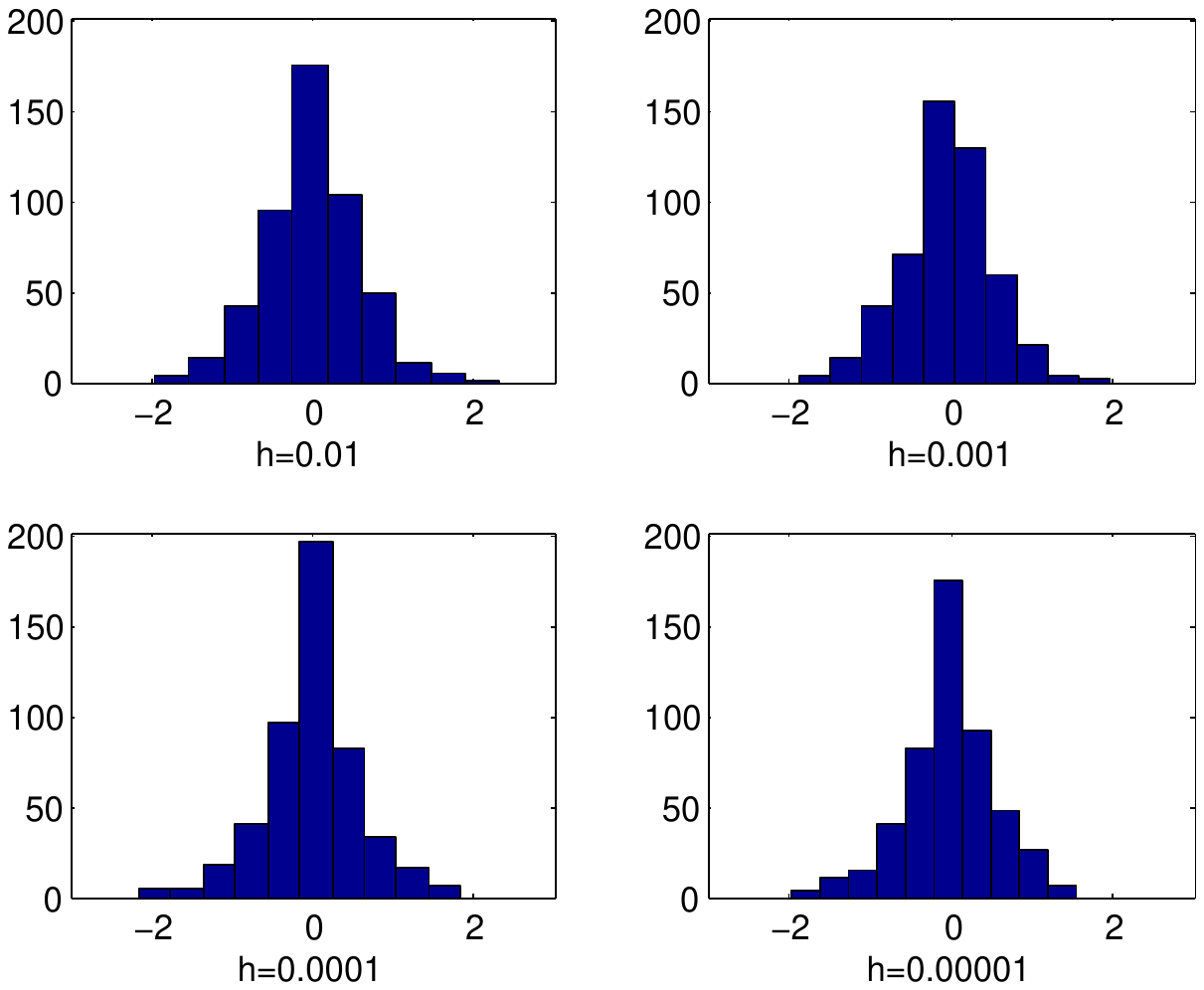}
\caption{Histograms of $500$ simulations of $x_T$ with different
  step-size.}
\label{fig:FlereSimuHforskellig}}
\qquad 
\parbox{2.6in}{
\includegraphics[trim=5cm 10cm 5cm 10cm,
width=0.40\textwidth]{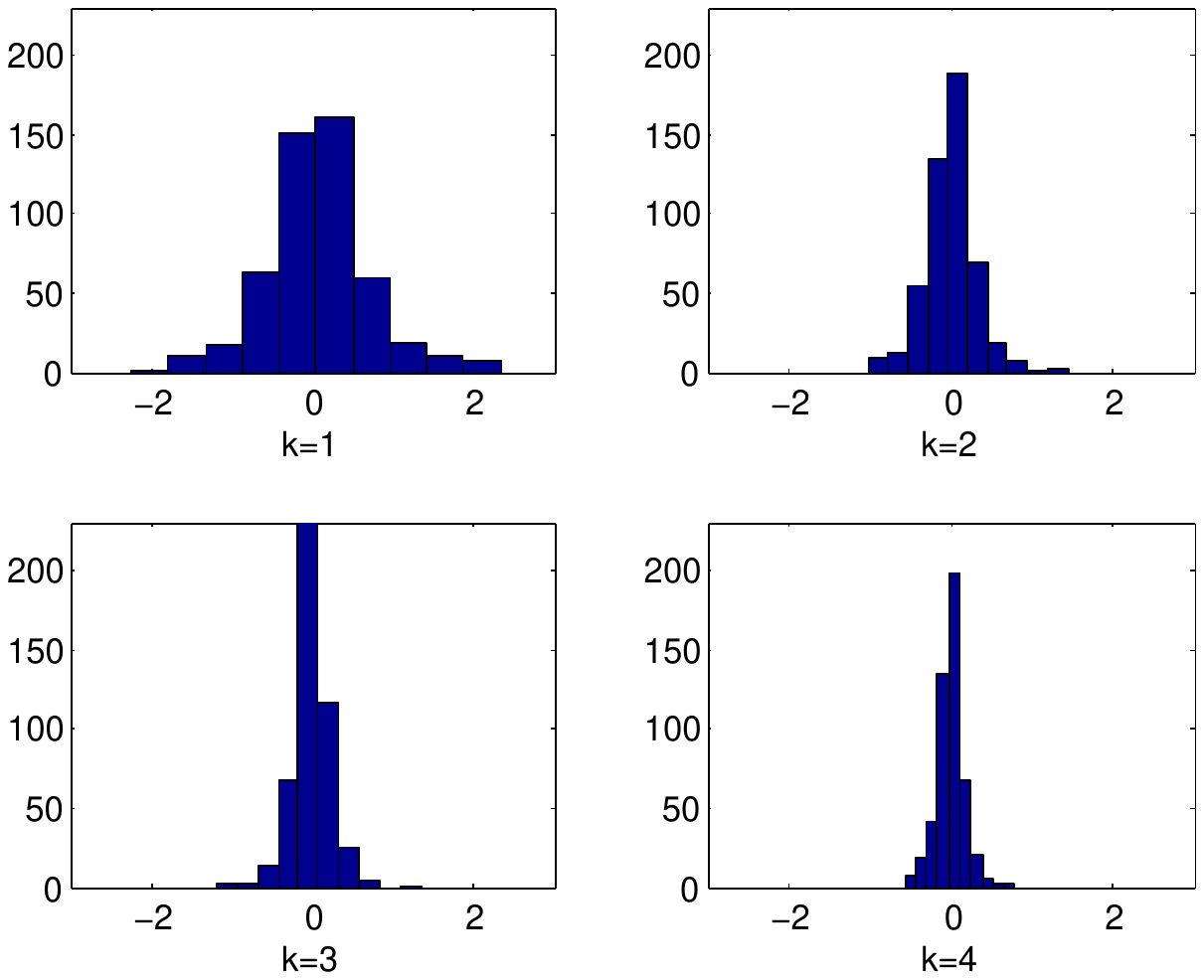}
\caption{Histograms of $500$ simulations of $x_T$ with different
  control gains. $h=0.001$.}
\label{fig:HistFlerKforskellig}}
\end{figure}

In order to investigate the influence of the step-size, figure
\ref{fig:FlereSimuHforskellig} illustrates four different histograms
of $500$ simulations. Here $h$ is $0.01,0.001,0.0001$ and $0.00001$
respectively and $k=1$. It can be seen right away that the result
narrows slightly around zero when $h$ becomes smaller, but changing in
the step-size does not immediately give big effect.

In figure \ref{fig:HistFlerKforskellig}, the control gain is changing,
$k=1,2,3,4$ and $h=0.001$. Here it is clear that changing  $k$
has an influence on the result of $x_T$. The variance of
$x_T$ gets smaller when $k$ increases. This is not surprising
since the overall influence of the random variable $W_n$ is decreased
when $k$ increases as mentioned in previous section.

\subsection{Theoretical Distribution of $x_n$} \label{sec:TheoDisxn}

In the following, we investigate 
a distribution of $x_n$ defined by the Euler-Maruyama method.

Consider the recursive determination of $x_{n+1}$ in \eqref{eq:deterministicStep}. Define an intermediate variable $z_n = x_n+hk\sign(x_n) $ and let $f_{z_n}$ and $f_n$ denote the density functions of $z_n$ and $x_n$, respectively. 
Moreover, let $N(0,h)$ denote the density function for $W_n$. 
From probability theory we get
\begin{equation}
 f_{n+1}(x) = f_{z_n}*N(0,h) \;,
\label{eq:convolutionDensityFunctions}
\end{equation}
where $*$ denotes convolution. 

Hence, we proceed by studying the density function $f_{z_n}(z)$. 
Let the distribution function of $z_n$ be denoted by $F_{z_n}$ such that $ F_{z_n}(z) = P(z_n\leq z)=P(x_n-hk\sign(x_n) \leq z)$, which can be expressed by 
\begin{eqnarray*}
P(z_n\leq z) &=& P(x_n-hk \leq z, x_n > 0) + P(x_n+hk \leq z ,x_n< 0)+ P(x_n \leq z ,x_n=0) \nonumber\\
 &=& P(x_n \leq z+hk, x_n > 0) + P(x_n \leq z-hk ,x_n< 0)   \;.
\label{eq:pzn}
\end{eqnarray*} 
For different values of $z$, the probability $P(z_n\leq z)$ can be expressed differently. If $z<-hk$ 
\begin{eqnarray*}
 P(z_n\leq z) &=& P(x_n \leq z-hk ,x_n< 0) = P(x_n \leq z-hk) =F_n(z-hk) \;.
\end{eqnarray*}
If $z>hk$ 
\begin{eqnarray*}
 P(z_n \leq z) 
 &=& P(x_n\leq z+hk,x_n>0) + P(x_n\leq z-hk) = P(x_n\leq z+hk) = F_n(z+hk)\;,
\end{eqnarray*}
and if $-hk\leq z \leq hk$
\begin{eqnarray*}
P(z_n \leq z)  &=& P(x_n \leq z+hk) - P(x_n< 0)   + P(x_n\leq z-hk) \\
 &=& F_n(z+hk)- F_n(0)+F_n(z-hk)\;.
\end{eqnarray*}
By introducing the indicator function $\mathbb{I}$, the expression of the distribution function of $z_n$ is
\begin{eqnarray*}
P(z_n \leq z) &=& F_n(z-hk) \mathbb{I}_{(-\infty,-hk)}(z)   + (F_n(z+hk)- F_n(0)+F_n(z-hk))\mathbb{I}_{[-hk, hk]}(z)  + F_n(z+hk) \mathbb{I}_{(hk,\infty)}(z) \\
 &=& F_n(z-hk) \mathbb{I}_{(-\infty,hk]}(z)   + F_n(z+hk) \mathbb{I}_{[-hk,\infty)}(z) - F_n(0) \mathbb{I}_{[-hk, hk]}(z)\;.
\end{eqnarray*}
By differentiating with respect to $z$, the density function of $z_n$ is  
\begin{eqnarray*}
\frac{\partial}{\partial z} F_{z_n} &=&  f_n(z-hk) \mathbb{I}_{(-\infty,hk]}(z)  - F_n(z-hk) \delta (z-hk) +f_n(z+hk)\mathbb{I}_{[-hk,\infty)}(z) \\ 
& & + F_n(z+hk) \delta(z+hk) - F_n(0)(\delta(z+hk)-\delta(z-hk))\\
&=& f_n(z-hk) \mathbb{I}_{(-\infty,hk]}(z) + f_n(z+hk)\mathbb{I}_{(-hk,\infty)}(z) + \delta(z+hk)(F_n(z+hk)-F_n(0)) \\
& & + \delta(z-hk)(F_n(0)-F_n(z-hk)) \\  
&=& f_n(z-hk) \mathbb{I}_{(-\infty,hk]}(z) + f_n(z+hk)\mathbb{I}_{[-hk,\infty)}(z) \;.
\end{eqnarray*} 
Therefore
\begin{equation}
  f_{z_n}(z) = f_n(z-hk) \mathbb{I}_{(-\infty,hk]}(z) + f_n(z+hk)\mathbb{I}_{[-hk,\infty)}(z) \;.
\label{eq:densityFunctionfzn}
\end{equation}
By substituting the above into \eqref{eq:convolutionDensityFunctions}, the density function $f_{n+1}$ is found from the density function $f_{n}$.
In the following section, the  solution to \eqref{eq:densityFunctionfzn} is investigated numerically. 

\subsection{Recursive Developing of the Density Function in Matlab} \label{RecursiveDensityFunctionSimulation}
The recursive density function is given by 
\begin{eqnarray}
f_{n+1}(x)&=& \left( f_n(x+kh) \mathbb{I}_{(-\infty,hk]} + f_n(x-hk) \mathbb{I}_{[-hk,\infty)} \right) *N(0,h) \;.
\label{eq:recursivn}
\end{eqnarray}
Following, we apply  Matlab to investigate the evolution of the function $f_{n+1}(x)$ for $n$ increasing.
Assume that the density function for the initial condition $x_0=c$ is  normal distributed with mean zero and variance $h$.
The Euler-Maruyama method is expected to converge to a stochastic process (or a distribution of a stochastic process) when $h\rightarrow 0$. (Under certain regularity conditions, so actually we cannot expect it here but only conjecture.)
We hope that the developing of the recursive density functions in \eqref{eq:recursivn} will reach stationary condition for $n\rightarrow \infty$. For this reason, the number $n$ of simulations is chosen to depend on the step size, such that $n = \left\lceil \frac{1}{h^{1+\alpha}} \right\rceil$, where $\alpha>0$. This ensures that both convergence criteria are fulfilled. 
 
\begin{figure}%
\centering
\includegraphics[trim=5cm 10cm 5cm 10cm, width=0.40\textwidth]{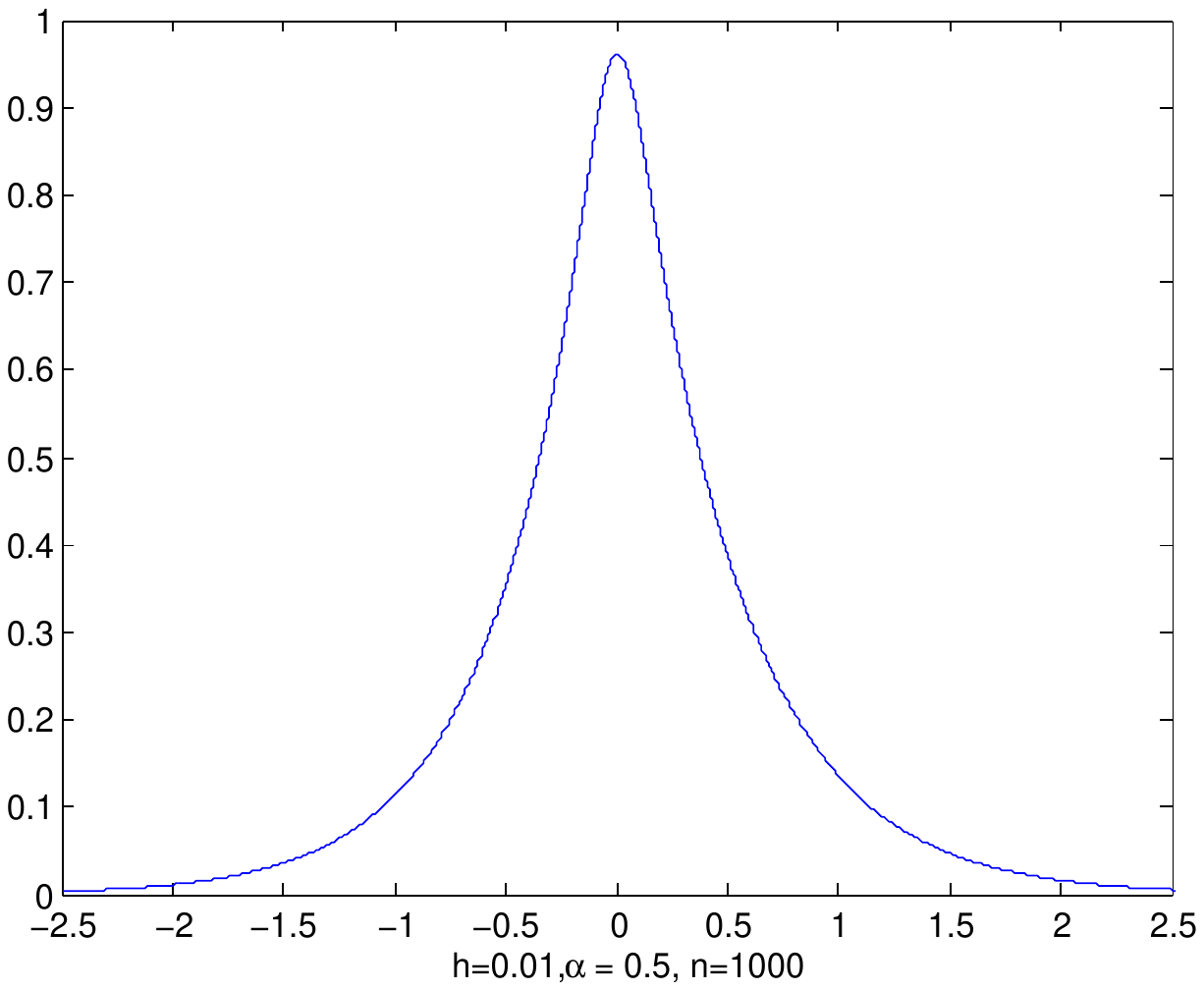}
	\caption{Recursive density function, $f_{1000}$.}
	\label{fig:recursiveDensity}
\end{figure}

Equation \eqref{eq:recursivn} is simulated in Matlab for $h=0.01, \alpha=0.5, k=1$ such that $n=1000$, the result is shown in figure \ref{fig:recursiveDensity}. At the end of section \ref{FokkerPlanck}, a comparison between the convergence of the recursive density function and the result obtained there is presented. 
   
In the following, we continue the study of \eqref{eq:densityFunctionfzn} under stationary assumptions. 

\subsection{Stationary State} \label{sec:StationaryState}

In the sequel we continue the study of the recursive density function under the assumptions that it is possible to reach stationary state in \eqref{eq:recursivn} for $n\rightarrow \infty$, such that 
\begin{equation*}
  f_{z_n}(z) = f(z-hk) \mathbb{I}_{(-\infty,hk]}(z) + f(z+hk)\mathbb{I}_{(-hk,\infty)}(z) \;,
\end{equation*}
where $f(x)$ is the stationary density function of the recursive development of $f_{n}(x)$. 
In this case we define an operator $H_h$ taking $f_n$ to $f_{n+1}$ by 
\begin{eqnarray*}
  H_h[f](x) &=& \int_{\mathbb{R}} \left( f(z-hk) \mathbb{I}_{(-\infty,hk]}(z)+ f(z+hk)\mathbb{I}_{(-hk,\infty)}(z) \right)  \frac{1}{\sqrt{2\pi h}}\exp\left(\frac{-(x-z)^2}{2h}\right) dz \;.
\end{eqnarray*} 
By taking the derivative with respect to $h$ and then the limit 
$h \rightarrow 0$,  we obtain (see appendix \ref{app:OperatorH}), 
\begin{equation}
 \lim_{h\rightarrow 0} \frac{\partial}{\partial h}H_h[f](x) = \left\{ \begin{array}{cll}
	 -kf'(x)+\frac{1}{2}f''(x) & \mbox{ for } & x < 0 \\
	  \psi(x) & \mbox{ for } & x =0 \\
	   kf'(x)+\frac{1}{2}f''(x)& \mbox{ for } & x > 0 
\end{array} \right. \;.
\label{eq:lignerFokker}
\end{equation}
The case $x=0$ is not important for the sequel developing, hence we leave $\psi(x)$ unspecified. Note that the right hand side of the first and last case in \eqref{eq:lignerFokker} have similarity with the stationary Fokker-Planck equation.

An approximation of the operator $H_h$ is 
\begin{eqnarray}
  H_h[f](x) &\approx& f(x) + h(k\sign(x)f'(x)+ \frac{1}{2}f''(x)) \nonumber \\
   &=& f(x) + hG[f](x) \;.
\label{eq:HnfapproxfhG}
\end{eqnarray}  
For a  fixed $h$ define $f_h$ to be the stationary density function and assume that  $\lim_{h\rightarrow 0} f_h$ exists, say $f_0$ and that $H_h(f_h) = f_h$. Hence, if we disregard the approximation in \eqref{eq:HnfapproxfhG}, we look for a function $f_h$ such that $G[f_h]=0$, which by \eqref{eq:lignerFokker} means that $f_h$ is a stationary solution to the Fokker-Planck equation. 
We conjecture that this heuristic will constitute the main ideas in the proof that the stationary distribution $f_0~(=\lim_{h\rightarrow 0}f_h)$ of  the Euler-Maruyama simulation converges to the Fokker-Planck equation. 


\section{Fokker-Planck Equation} \label{FokkerPlanck}
Based on the preceding work, this section introduces the one-dimensional Fokker-Planck
equation and apply this to determine a density function of $x_t$ for fixed $t$.

Let $x_t$ be a solution to \eqref{eq:basisdiff}. For a fixed
$t\in[0,T]$ let $\phi(x,t)$ be the density function of
$x_t$, such that
\begin{equation}
\int_{\mathbb{R}}{\phi(x,t)} dx= 1
\label{eq:bound}
\end{equation}
with initial condition
\begin{equation*}
\lim_{t\rightarrow 0} \phi(x,t) = \delta(x-x_0)\;,
\end{equation*}
where $\delta$ denotes the Dirac-delta function.
If the drift and diffusion coefficient are moderately smooth functions,
then the density function $\phi(x,t)$ satisfies the Fokker-Plank
equation \cite{Fokker,Arnold}. For \eqref{eq:basisdiff} this means
\begin{equation}
\frac{\partial}{\partial t} \phi(x,t) = \frac{\partial}{\partial x} k
\sign(x) \phi(x,t) + \frac{1}{2} \frac{\partial^2}{\partial x^2}
\phi(x,t) \;.
\label{eq:FokkerPlanckEq}
\end{equation}
As mentioned previous, the drift coefficient $-k \sign(x)$ is not a
smooth function and for this reason there is no guarantee that
\eqref{eq:FokkerPlanckEq} is valid.

\subsection{Solution in Two Domains} \label{sec:FokkerToDomain}
\begin{figure}%
\centering
\parbox{2.6in}{
\includegraphics[trim=5cm 10cm 5cm 10cm,
width=0.40\textwidth]{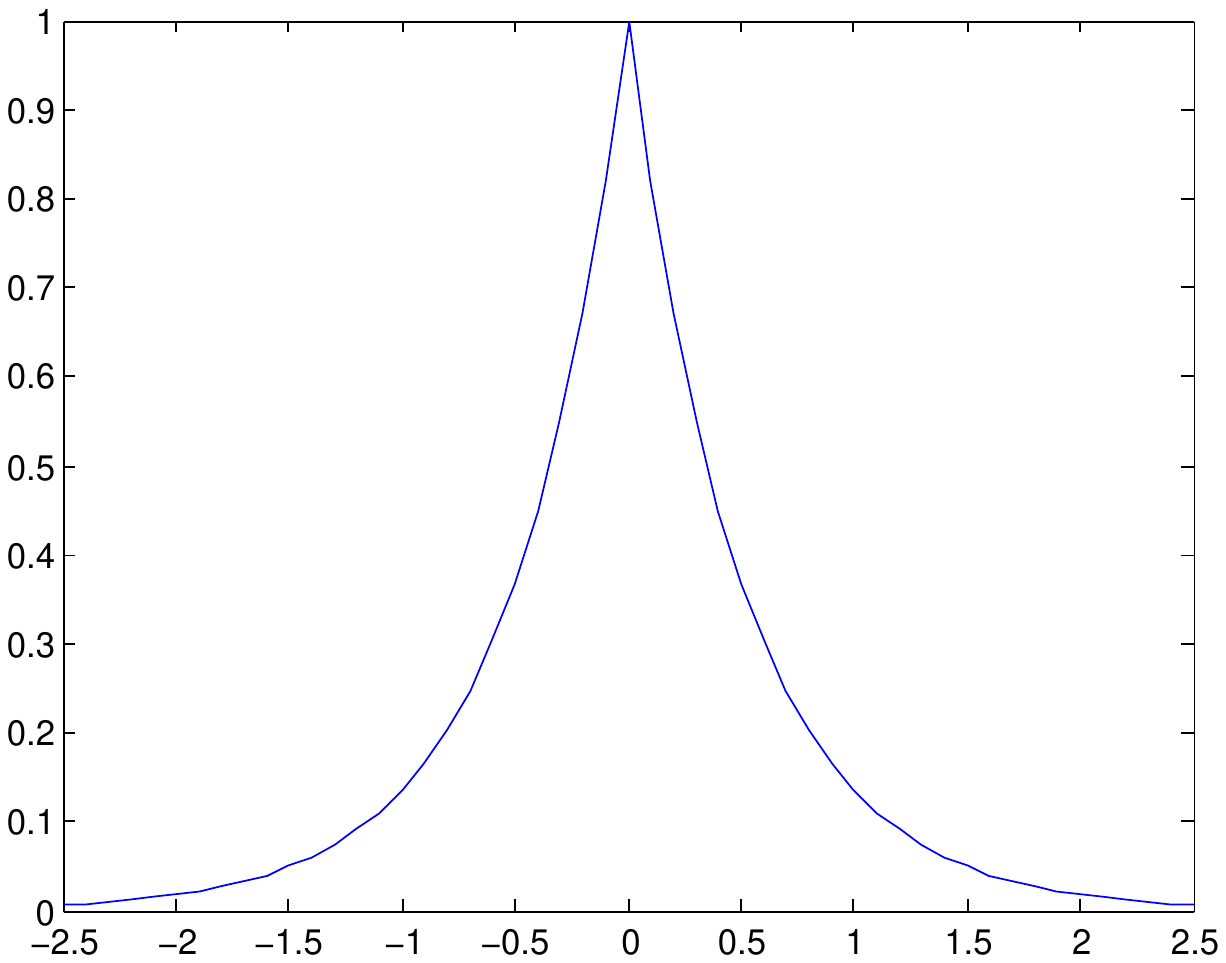}
\caption{Density function $\phi(x,t)$ for $x_t$ for fixed
  $t\in[0,T]$.}
\label{fig:DensityFunctionFokkerPlanckdomianSolution}}
\qquad 
\parbox{2.6in}{%
\includegraphics[trim=5cm 10cm 5cm 10cm, width=0.40\textwidth]{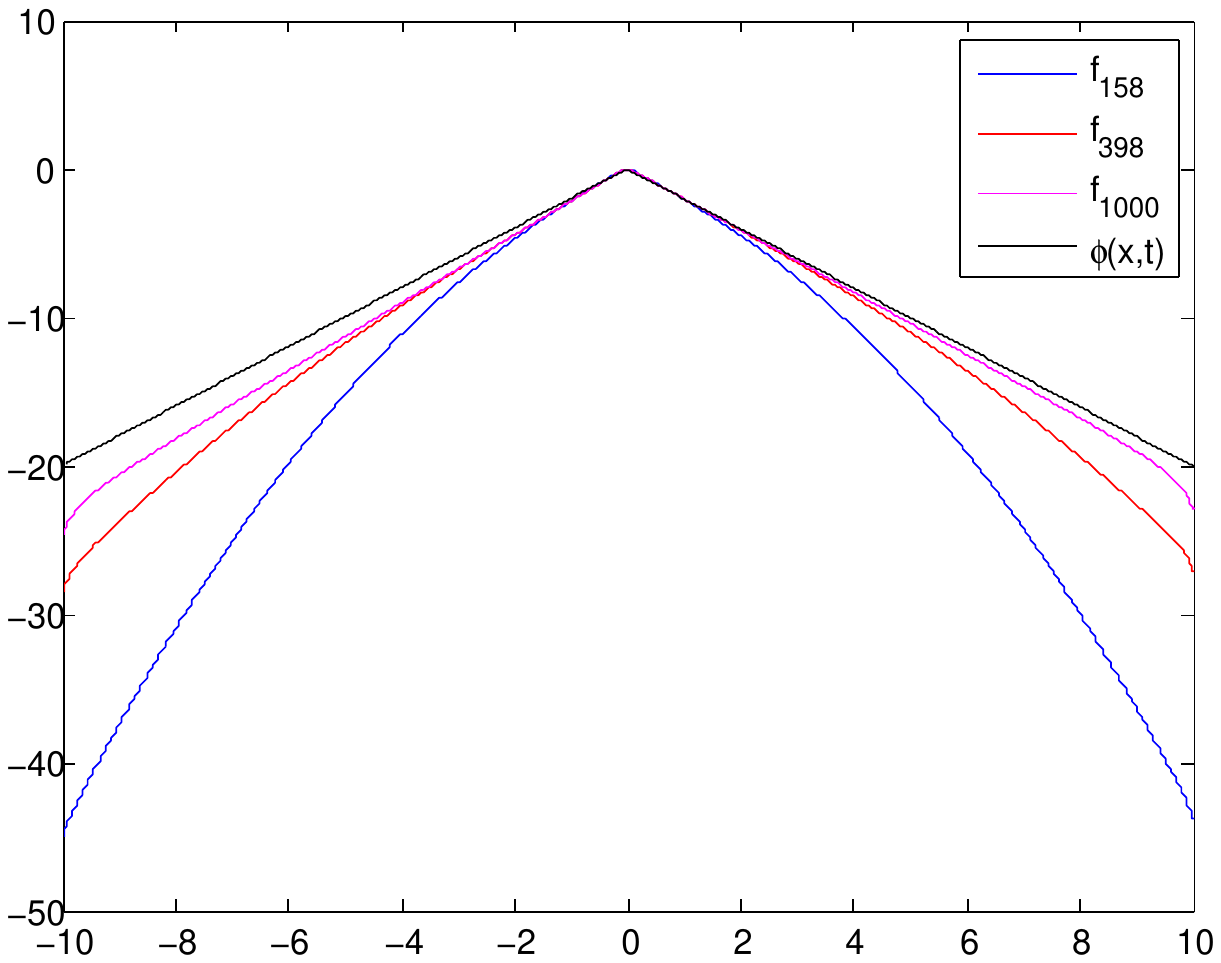}
	\caption{Logarithmic stationary density function and logarithmic recursive density function of $f_{158},f_{398},f_{1000}$.}
	\label{fig:recursiveLogDensityComparison}}
\end{figure}

In order to avoid the discontinuous challenges by the $\sign$-function, we consider the Fokker-Planck equation \eqref{eq:FokkerPlanckEq} in the
domains $(-\infty,0)$ and $(0,\infty)$. 
This gives
\begin{eqnarray*}
\frac{\partial}{\partial t} \phi(x,t) &=& \frac{\partial}{\partial x}
k  \phi(x,t) + \frac{1}{2} \frac{\partial^2}{\partial x^2} \phi(x,t)
\mbox{ for } x>0 \\ 
\frac{\partial}{\partial t} \phi(x,t) &=& -\frac{\partial}{\partial x}
k  \phi(x,t) + \frac{1}{2} \frac{\partial^2}{\partial x^2} \phi(x,t)
\mbox{ for } x<0 \;. 
\end{eqnarray*}
Assume that the density function can reach stationarity, then
\begin{eqnarray*}
0&=& \frac{\partial}{\partial x} k  \phi(x,t)  + \frac{1}{2}
\frac{\partial^2}{\partial x^2} \phi(x,t) \mbox{ for } x>0 \\ 
0&=& -\frac{\partial}{\partial x} k  \phi(x,t) + \frac{1}{2}
\frac{\partial^2}{\partial x^2} \phi(x,t) \mbox{ for } x<0 \;, 
\end{eqnarray*}
which are two ODEs.  The characteristic
equations are
\begin{eqnarray*}
ks+ \frac{1}{2} s^2 &=& 0 \mbox{ for } x>0 \\
-ks+ \frac{1}{2} s^2&=&0 \mbox{ for } x<0 \;,
\end{eqnarray*}
which have the roots
\begin{eqnarray*}
s = 0, & & s=- 2k \quad \mbox{ for } x>0 \\
s = 0, & & s = 2k \quad \mbox{ for } x<0 \;,
\end{eqnarray*}
such that
\begin{eqnarray*}
\phi_+(x) &=& c_1 \exp(0x) + d_1 \exp(-2kx) \mbox{ for } x>0 \\
\phi_-(x) &=& c_2 \exp(0 x) + d_2\exp(2kx) \mbox{ for } x<0
\end{eqnarray*}
are solutions of \eqref{eq:FokkerPlanckEq} in the respective domain.
We discard the constant term and, due to symmetry around $x=0$, it can be expected that $d_1=d_2:=d$.  The boundary
constraint in equation \eqref{eq:bound} gives
\begin{equation*}
d = \frac{1}{2 \int_0^\infty{\exp(-2kx)}} = k \;.
\end{equation*} 
Furthermore, from the above we assume that $\phi(x,t)=k$ for $x=0$.
This gives the density function
\begin{equation}
\phi(x,t) = \left\{ \begin{array}{cl}
k \exp(-2kx) & \mbox{ for } x>0 \\
k\exp(2kx) & \mbox{ for } x<0 \\
k & \mbox { for } x=0 
\end{array} \right.
 \label{eq:FokkerDensity}
\end{equation}
which is illustrated in figure
\ref{fig:DensityFunctionFokkerPlanckdomianSolution} for $k=1$.
The density function $\phi(x,t)$ is the Laplace distribution with location parameter zero and scale parameter $\frac{1}{2k}$. 

The density function in \eqref{eq:FokkerDensity} is compared with the developing of the recursive density functions from section \ref{RecursiveDensityFunctionSimulation} by considering the logarithmic of the density functions.   
The recursive cases with $n=158,398,1000$ are shown in figure \ref{fig:recursiveLogDensityComparison} together with the logarithmic stationary density function $\log\left(\phi(x,t)\right)$.
For increasing $n$, the logarithmic recursive density function gets closer to the logarithmic stationary density function, as we hope to observe.

Due to the lack of continuity in all the previous investigations, in the following section we construct a smooth function which approximate the $\sign$-function and investigate which results the applied methods provide with this smooth function as input.   


\section{Approximation of sign-Function} \label{sec:Approxsign}
\begin{figure}
\centering
\includegraphics[trim=5cm 10cm 5cm 10cm,
width=0.4\textwidth]{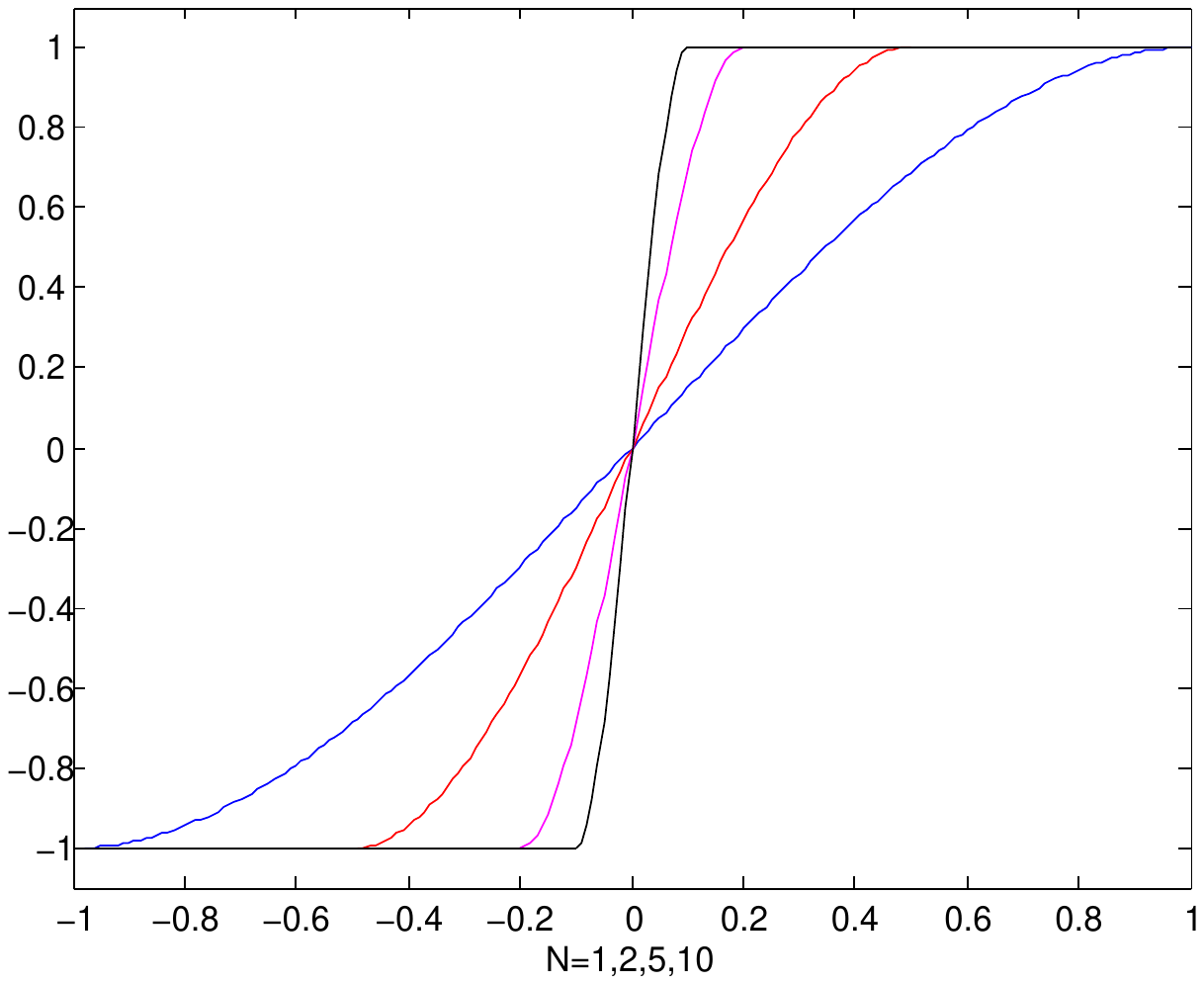}
\caption{Approximation of the $\sign$-function.}
\label{fig:ApproxSignFunction}
\end{figure}

As mentioned, the problem with the $\sign$-function is that it is not
smooth. 
For this reason we construct a function $f_N(x)$ which has pointwise convergence to the $\sign$-function as $N\rightarrow \infty$.
\begin{eqnarray}
f_N(x) = \left\{ \begin{array}{cl}
1 & \mbox{ for } x>\frac{1}{N} \\
-\frac{N^3}{2}x^3+\frac{3N}{2}x & \mbox{ for }  - \frac{1}{N} \leq x \leq \frac{1}{N} \\
-1 & \mbox{ for } x<-\frac{1}{N} 
\end{array} \right. \;.
\label{eq:ApproxSignFuncktion}
\end{eqnarray}
 In figure \ref{fig:ApproxSignFunction}, the functions $f_1(x),f_2(x),f_5(x)$ and
$f_{10}(x)$ are shown.
In the following, we will apply the Euler-Maruyama method to this function.

\subsection{Euler-Maruyama Method Applied to the Approximated  sign-Function}
\begin{figure}%
\centering
\parbox{2.6in}{%
\includegraphics[trim=5cm 10cm 5cm 10cm,
width=0.42\textwidth]{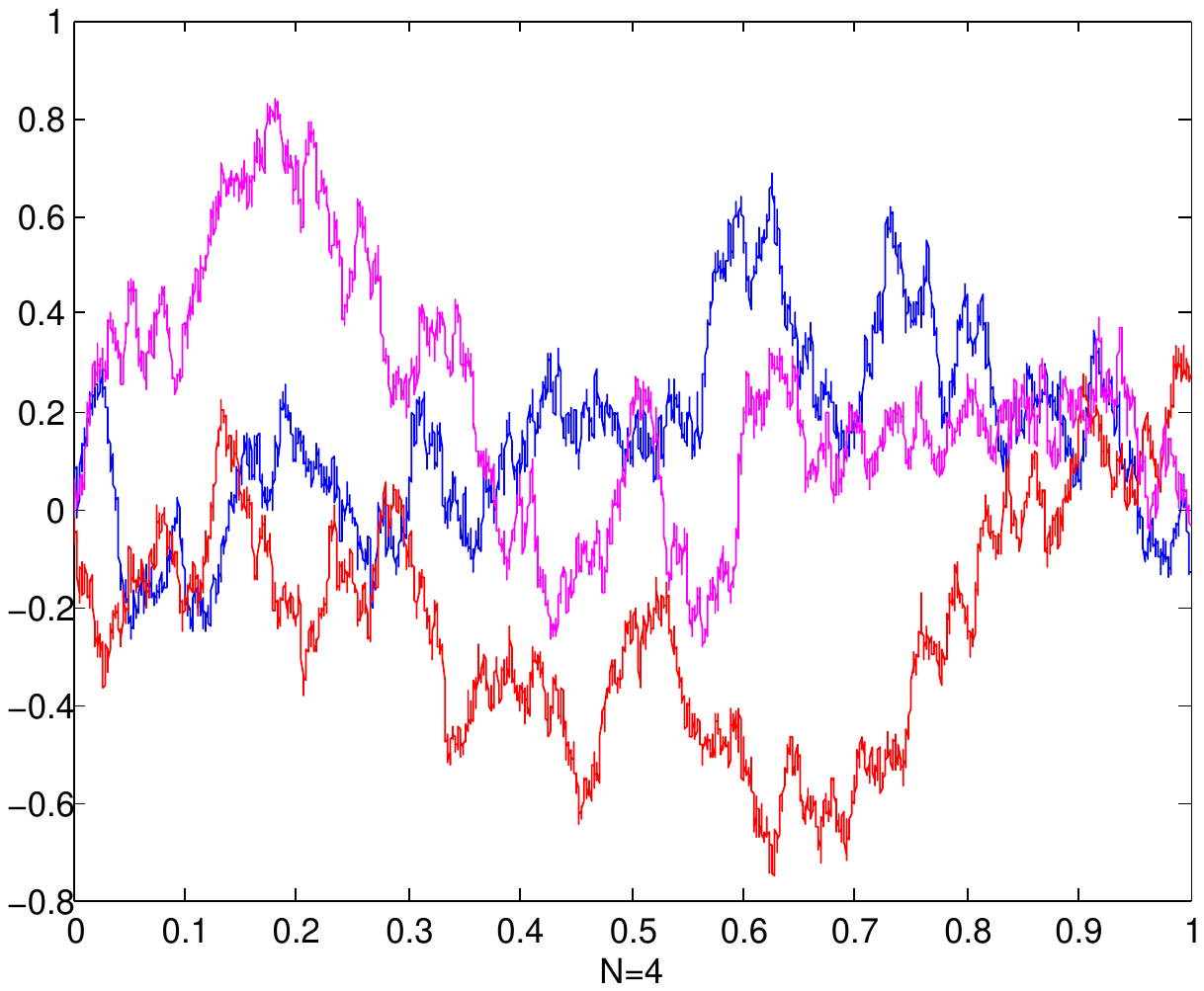}
\caption{Realizations of $x_t$ to a SDE with drift $b(t,x_t)=-kf_4(x_t)$.}
\label{fig:3approxXtforapproxSign}}
\qquad \qquad 
\parbox{2.6in}{
\includegraphics[trim=5cm 10cm 5cm 10cm,
width=0.42\textwidth]{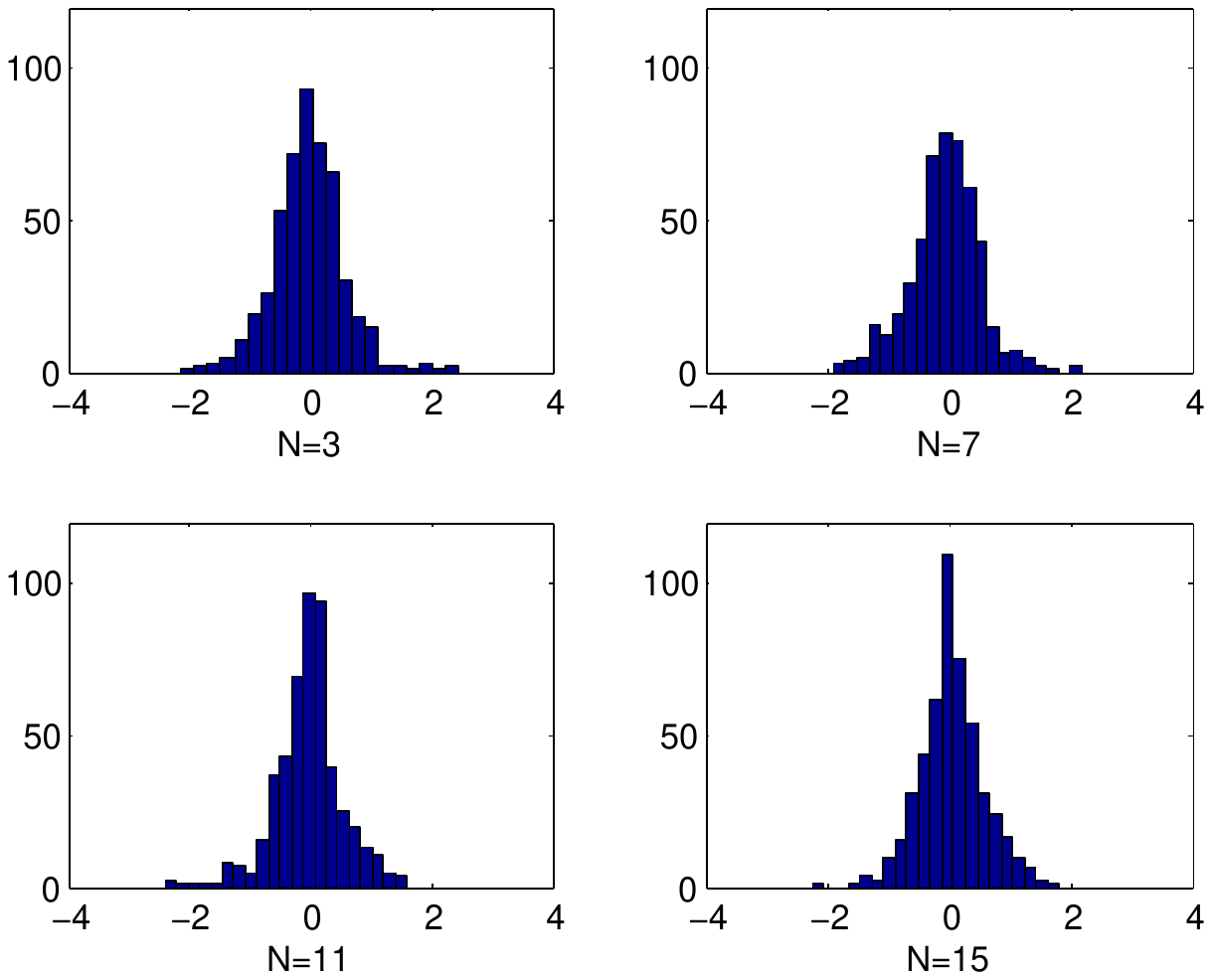}
\caption{Histogram of $500$ simulations of $x_T$ where different
  approximations to the $\sign$-function are used.}
\label{fig:4histApproxSignforskellign}}
\end{figure}

It is proved that the Euler-Maruyama method exhibits convergence to the solution of the SDE if the drift and diffusion coefficient satisfy certain regularity conditions. Since $f_N(x)$ from \eqref{eq:ApproxSignFuncktion} is a smooth function, we apply the Euler-Maruyama method with this input instead of the $\sign$-function.  
 The same procedure as in section \ref{simulation} is applied, such that
\begin{equation}
x_{j+1} = x_j - h k f_N(x_j)+W_j \;.
\label{eq:deterministicStepApproxSign}
\end{equation}
Since the slope of $f_N(x)$ increases significantly with the increase of $N$, the step-size is chosen to be dependent on $N$, such that $h = \frac{0.001}{N}$. We hope to avoid inaccurate simulations due to the step-size with this method. (If $x_{i}\in[-\frac{1}{N},\frac{1}{N}]$ for $t_i\in[0,T]$ and $N$ is big, then the value of $f_N(x_i)$ is dominating in the determination of $x_{i+1}$ such that $x_{i}<<x_{i+1}$ or $x_i >> x_{i+1}$.) 
 As previous, $k=1$.
In figure \ref{fig:3approxXtforapproxSign}, are shown $3$
realizations of $x_t$ when $f_4(x)$ is used to approximate $\sign(x)$.
In figure \ref{fig:4histApproxSignforskellign} histograms of $500$
simulations are presented based on different approximations to the
$\sign$-function.

In the following section, the smooth function is implemented in the Fokker-Planck equation.

\subsection{Solution to the Fokker-Planck Equation with Approximated
  sign-function} \label{SolutionFokPlanckAproxSign} 
Previous, the Fokker-Planck equation have been applied to a SDE where the drift is discontinuous due to the $\sign$-function. As mentioned, we cannot expect this to be meaningful. Following the smooth function $f_N(x)$ is used as a substitute for the $\sign$-function, such that we consider the SDE 
\begin{eqnarray}
  dx_t = -kf_N(x_t)dt+dB_t \;.
\label{eq:SDEapproxSign}
\end{eqnarray}
We investigate if it is possible to determine a density function that is a solution to the Fokker-Planck equation with $-kf_N(x)$ as the drift. 
    
\begin{figure}
	\centering
		\includegraphics[trim=5cm 10cm 5cm 10cm, width=0.40\textwidth]{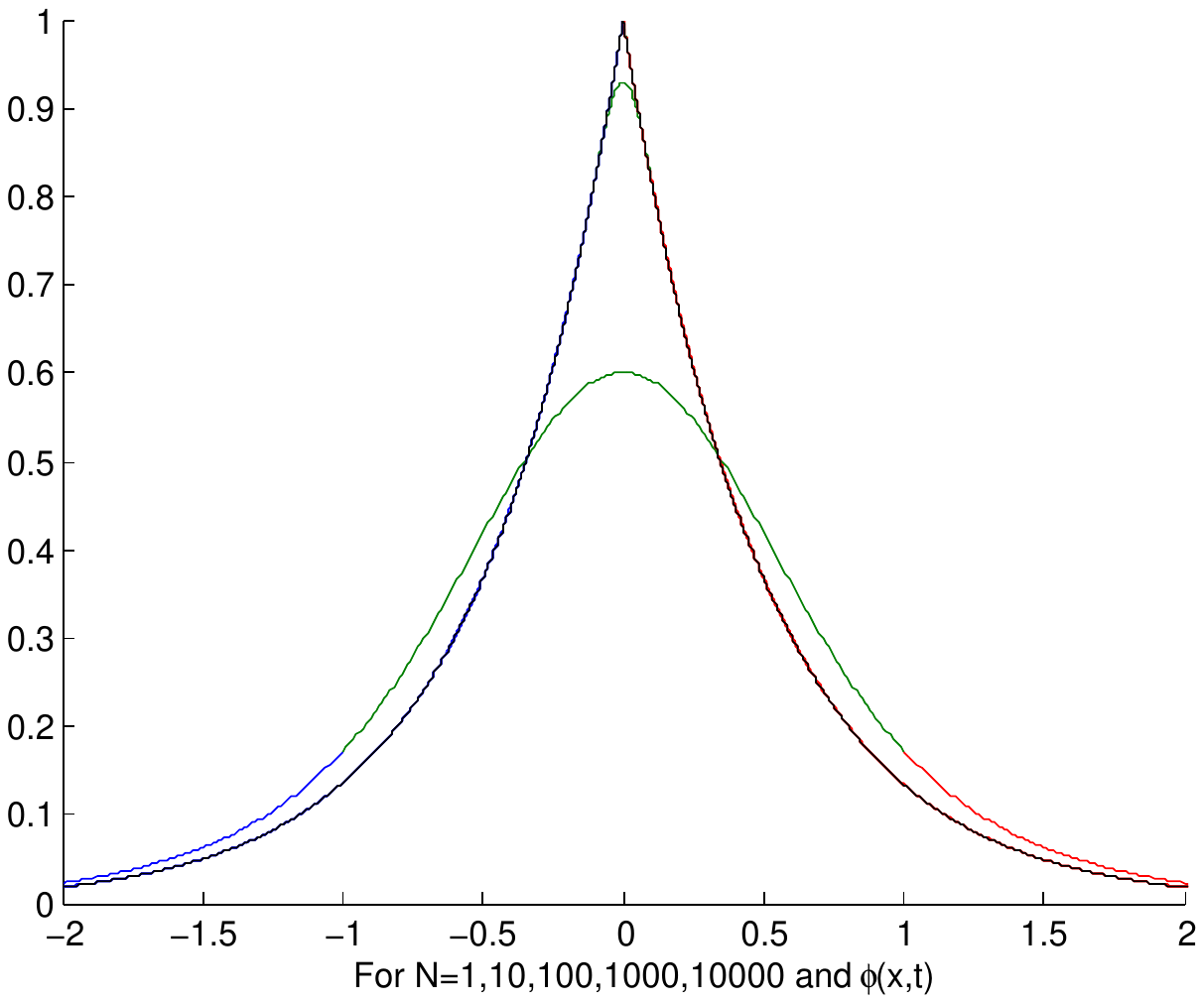}
	\caption{Density functions for solutions to \eqref{eq:SDEapproxSign} together with $\phi(x,t)$.}
	\label{fig:ApproxSignFokkerPlanckEq}
\end{figure}

With same technique as previous we consider the Fokker-Planck equation in three domains. First the domain
$[-\frac{1}{N}, \frac{1}{N}]$ is considered with $f_N(x)$ as an approximation of
$\sign(x)$.
\begin{eqnarray}
\frac{\partial}{\partial t} \phi_N(x,t) &=& \frac{\partial}{\partial x}
k \left(-\frac{N^3}{2}x^3+\frac{3N}{2}x \right)  \phi_N(x,t)  + \frac{1}{2} \frac{\partial^2}{\partial x^2} \phi_N(x,t) \;.
\label{eq:fokkerlimit}
\end{eqnarray} 
Under stationary assumptions, the function $\phi_N(x,t)= \phi_0\exp\left( \frac{N^3}{4}kx^4 - \frac{3N}{2}kx^2\right)$ fulfills \eqref{eq:fokkerlimit}. By including the domains $(-\infty,-\frac{1}{N})$ and $(\frac{1}{N},\infty)$, the
density function for solutions to  \eqref{eq:SDEapproxSign}
becomes
\begin{eqnarray*}
\phi_N(x,t) = \left\{ \begin{array}{cl}
d \exp(2kx) & \mbox{ for } x < - \frac{1}{N} \\
\phi_0\exp\left( \frac{N^3}{4}kx^4 - \frac{3N}{2}kx^2\right) & \mbox{ for } -\frac{1}{N} \leq x \leq
\frac{1}{N} \\ 
d \exp(-2kx) & \mbox{ for } x > \frac{1}{N}
\end{array} \right.\;,
\label{eq:phin}
\end{eqnarray*} 
with constraints
\begin{eqnarray*}
1 &=& \int_{-\infty}^{-\frac{1}{N}} d \exp(2kx) dx +  \int_{\frac{1}{N}}^\infty d \exp(-2kx) dx  + \int_{-\frac{1}{N}}^\frac{1}{N} \phi_0
\exp\left(k\left(\frac{N^3}{4}x^4- \frac{3N}{2}x^2\right)\right) dx  
\end{eqnarray*}
and
\begin{equation}
\lim_{x \rightarrow \pm\frac{1}{N}+ }\phi_N(x,t) = \lim_{x \rightarrow \pm \frac{1}{N}-} \phi_N(x,t) \;.
 \label{eq:constraints}
\end{equation}
From \eqref{eq:constraints} 
\begin{eqnarray*}
d &=& \phi_0 \exp\left(\frac{3k}{4N} \right) \;,
\end{eqnarray*}
so the normalization constant becomes
\begin{eqnarray*}
  \phi_0 = \frac{1}{2 \int_0^{\frac{1}{N}} \exp\left( -\frac{3kN}{2}x^2 + \frac{kN^3}{4}x^4\right) dx + \frac{1}{k} \exp\left(-\frac{5k}{4N}\right)}
\end{eqnarray*}
and 
\begin{eqnarray*}
\phi_N(x,t) &=& \left\{ \begin{array}{cl}
\phi_0 \exp\left( \frac{3k}{4N} \right) \exp(2kx) & \mbox{ for } x < - \frac{1}{N} \\
\phi_0\exp\left( \frac{N^3}{4}kx^4 - \frac{3N}{2}kx^2\right) & \mbox{ for } -\frac{1}{N} \leq x \leq
\frac{1}{N} \\ 
\phi_0 \exp\left( \frac{3k}{4N} \right) \exp(-2kx) & \mbox{ for } x > \frac{1}{N}
\end{array} \right.\;.
\end{eqnarray*} 
This density function is simulated in Matlab with $\phi_0$ numerically calculated. In figure \ref{fig:ApproxSignFokkerPlanckEq} results  are shown for $N=1,10,100,1000,10000$ together with the stationary density function from the Fokker-Planck equation $\phi(x,t)$. Note that $\phi_{100},\phi_{1000}$ and $\phi_{10000}$ are not visible. 
The simulations show that $\phi_N(x,t) \rightarrow \phi(x,t)$ as expected.


\section{Fourier Transformation} \label{FourierTrans}
In section \ref{sec:TheoDisxn} we investigate a candidate density function for the intermediate variable $z_n$ in order to determine a density function for the stochastic process produced by Euler-Maruyama method. Another method than the one presented there is to Fourier transform the density function for $z_n$ before the convolution of the density functions is done.  
The Fourier transformation of \eqref{eq:densityFunctionfzn} is given by
\begin{eqnarray*}
\hat{F}(\omega) &=& \int_{\mathbb{R}} \left[ f_n(z-hk) \mathbb{I}_{(-\infty,hk]}  + f_n(z+hk)\mathbb{I}_{(-hk,\infty)} \right] \exp(-j\omega z) dz \;.
\end{eqnarray*}
By expansion and including of the Laplace transformation, see appendix \ref{app:FourierTrans}, the following relation appears: 
\begin{equation*}
\hat{F}(\omega) =  \cos(\omega hk) \real\left(\hat{F}(\omega)\right) - 2  \sin(\omega hk) \imag\left(\tilde{F}(\omega)\right) \;, 
\end{equation*}
where $\tilde{F}(\omega)$ is the Laplace transformation. 
For the moment we have no further interpretation of how this can help the developing of the recursive density function.

\section{Discussion}

This paper has presented initial studies of SDEs with discontinuous drift. 
Theoretical and numerical approaches have been applied to a particular SDE in order to investigate meaningful results in the form of density functions.   

The candidate recursive density function developed from the Euler-Maruyama method has the tendency to approximate the density function of the stationary Fokker-Planck equation, which strengthens the conjecture that these density functions actually exist.   
Furthermore, the fact that the Laplace distribution is a solution to the stationary Fokker-Planck equation with discontinuous drift supports the assumption that it is possible to give a meaningful answering to questions about definition  of solutions to discontinuous SDEs and their probabilistic properties. 
However, we have not completed these processes, yet. 

All in all, the different approaches indicate a connection between the candidate density function for the solution to the particular SDE with discontinuous drift and the Fokker-Planck equation. Though, the stationary assumptions have extensive impact on this result. Without this condition, the solvability of the Fokker-Planck equation would decrease significantly.

An immediate object for future studies is to formalize the heuristic presentation in this paper to obtain operational definition and  regular results connected to classes of SDEs with discontinuous drift.

\appendix
\section{Appendix}

\subsection{Approximation of the operator $H_h$} \label{app:OperatorH}
In section \ref{sec:StationaryState} the operator $H_h$ was defined by
\begin{eqnarray*}
  H_h[f](x) &=& \int_{\mathbb{R}} \left( f(z-hk) \mathbb{I}_{(-\infty,hk]}(z)+ f(z+hk)\mathbb{I}_{(-hk,\infty)}(z) \right)  \frac{1}{\sqrt{2\pi h}}\exp\left(\frac{-(x-z)^2}{2h}\right) dz \;,
\end{eqnarray*} 
Following, a notation $N_h(z) = \frac{1}{\sqrt{2\pi h}}\exp\left(\frac{-z^2}{2h}\right)$ is used and we consider the generator $\frac{\partial}{\partial h}H_h$. 
\begin{eqnarray}
\frac{\partial}{\partial h}H_h[f](x) &=& \int_{\mathbb{R}} \left( -kf'(z-hk) \mathbb{I}_{(-\infty,hk]}(z) + kf'(z+hk)\mathbb{I}_{[-hk,\infty)}(z) \right)  \cdot N_h(x-z) dz \nonumber\\
  & & +\int_{\mathbb{R}} \left( -kf(z-hk) \delta(z-hk)  +k f(z+hk)\delta(z+hk) \right) \cdot N_h(x-z) dz \nonumber\\
  & & + \int_{\mathbb{R}} \left( f(z-hk) \mathbb{I}_{(-\infty,hk]}(z) + f(z+hk)\mathbb{I}_{[-hk,\infty)}(z) \right) \frac{\partial}{\partial h}(N_h(x-z))dz \label{eq:lastterm} \;.
\end{eqnarray}
The second integral term above gives
\begin{eqnarray*}
  && \int_{\mathbb{R}} \left( -kf(z-hk) \delta(z-hk) +k f(z+hk)\delta(z+hk) \right) \cdot N_h(x-z) dz \\
  &=& kf(0)\left(N_h(x+hk)-N_h(x-hk)  \right) \;.
\end{eqnarray*}
The derivative of the last integral term in \eqref{eq:lastterm} gives 
\begin{eqnarray*}
  \frac{\partial}{\partial h}(N_h(x-z)) = \frac{1}{\sqrt{2\pi h}}\left( \frac{(x-z)^2}{2h^2}- \frac{1}{2h} \right) \exp\left(\frac{-(x-z)^2}{2h}\right) \;.
\end{eqnarray*}
Consider the derivative with respect to $z$,
\begin{eqnarray*}
  \frac{\partial}{\partial z}(N_h(x-z)) = \frac{1}{\sqrt{2\pi h}}\frac{-(x-z)}{h}\exp\left(\frac{-(x-z)^2}{2h}\right) \;,
\end{eqnarray*}
and the second derivative with respect to $z$,
\begin{eqnarray*}
 \frac{\partial^2}{\partial z^2}(N_h(x-z))= \frac{1}{\sqrt{2\pi h}}\left(  \frac{(x-z)^2}{h^2} + \frac{1}{h} \right)\exp\left(\frac{-(x-z)^2}{2h}\right) \;.
\end{eqnarray*}
Thus, for small $h$ 
\begin{eqnarray*}
  \frac{\partial}{\partial h}(N_h(x-z)) \approx \frac{1}{2} \frac{\partial^2}{\partial z^2}(N_h(x-z)) \;.
\end{eqnarray*}
Therefore, the last integral term in $\frac{\partial}{\partial h}H_h[f](x)$ is approximately
\begin{eqnarray*}
  & & \frac{1}{2} \int_{\mathbb{R}} \left( f(z-hk) \mathbb{I}_{(-\infty,hk]}(z) + f(z+hk)\mathbb{I}_{[-hk,\infty)}(z) \right)  \frac{\partial^2}{\partial z^2}(N_h(x-z))dz \\
  &=& \frac{1}{2} \int_{-\infty}^{hk} f(z-hk) \frac{\partial^2}{\partial z^2}(N_h(x-z))dz  +\frac{1}{2} \int^{\infty}_{-hk} f(z+hk) \frac{\partial^2}{\partial z^2}(N_h(x-z))dz \\
  &=&  \frac{1}{2} \left( [ f(z-hk) \frac{\partial}{\partial z}(N_h(x-z)) ]^{hk}_{-\infty}   - [f'(z-hk) N_h(x-z)]^{hk}_{-\infty} + \int_{-\infty}^{hk} f''(z-hk)N_h(x-z)dz \right)\\
  & & + \frac{1}{2} \left( [ f(z+hk) \frac{\partial}{\partial z}(N_h(x-z)) ]^{\infty}_{-hk}   - [f'(z+hk) N_h(x-z)]^{\infty}_{-hk}  + \int^{\infty}_{-hk} f''(z+hk)N_h(x-z)dz \right) \;.
\end{eqnarray*}
The limit of $\frac{\partial}{\partial h}H_h$ for $h \rightarrow 0$ is then 
\begin{equation*}
 \lim_{h\rightarrow 0} \frac{\partial}{\partial h}H_h[f](x) = \left\{ \begin{array}{lll}
	 -kf'(x)+\frac{1}{2}f''(x) & \mbox{ for } & x < 0 \\
	 \psi(x) & \mbox{ for } & x =0 \\
	   kf'(x)+\frac{1}{2}f''(x)& \mbox{ for } & x > 0 
\end{array} \right. \;,
\end{equation*}
where $\psi(x)$ is unknown. 

\subsection{Fourier Transformation Expansion} \label{app:FourierTrans}
In section \ref{FourierTrans} the Fourier transformation of $f_{z_n}$ is presented. Below follows the expansion. 
\begin{eqnarray*}
\hat{F}(\omega) &=& \int_{\mathbb{R}} \left[ f_n(z-hk) \mathbb{I}_{(-\infty,hk]}  + f_n(z+hk)\mathbb{I}_{(-hk,\infty)} \right] \exp(-j\omega z) dz \\
 &=& \int_{-\infty}^{hk}  f_n(z-hk)\exp(-j\omega z) dz  + \int_{-hk}^\infty f_n(z+hk)\exp(-j\omega z) dz \;.
\end{eqnarray*}
By changing variable  $x=z-hk$ in the first integral and $x=z+hk$ in the second integral, 
\begin{eqnarray}
 \hat{F}(\omega) &=&  \int_{-\infty}^{0}  f_n(x)\exp(-j\omega (x+hk)) dx + \int_{0}^\infty f_n(x)\exp(-j\omega (x-hk)) dx \nonumber \\
 &=& \exp(-j\omega hk))\int_{-\infty}^{0}  f_n(x)\exp(-j\omega x) dx + \exp(j\omega hk) \int_{0}^\infty f_n(x)\exp(-j\omega x) dx \nonumber \\
 &=& (\cos(\omega hk)-j\sin(\omega hk)) \int_{-\infty}^{0}  f_n(x)\exp(-j\omega x) dx  \nonumber \\
 & & \quad + (\cos(\omega hk)+j\sin(\omega hk)) \int_{0}^\infty f_n(x)\exp(-j\omega x) dx \label{eq:betragtreelpart} \;.
\end{eqnarray}
Based on the uncertainty around zero in the recursive determination of $x_{n+1}$ in \eqref{eq:deterministiskStep32}, we find it fair to assume that $f_n$ is an even function.  
Fourier transformations of even functions give zero in the imaginary part so only the real part of \eqref{eq:betragtreelpart} is considered. 
\begin{eqnarray}
\hat{F}(\omega) &=&  \cos(\omega hk) \real\left( \int_{-\infty}^{0}  f_n(x)\exp(-j\omega x) dx  \right) + \sin(\omega hk) \imag \left( \int_{-\infty}^{0}  f_n(x)\exp(-j\omega x) dx  \right) \nonumber\\
&& + \cos(\omega hk) \real\left( \int_{0}^\infty f_n(x)\exp(-j\omega x) dx \right)  - \sin(\omega hk) \imag\left(\int_{0}^\infty f_n(x)\exp(-j\omega x) dx \right)\nonumber \\
&=& \cos(\omega hk) \int_{-\infty}^{0}  f_n(x)\cos(\omega x) dx  - \sin(\omega hk) \int_{-\infty}^{0}  f_n(x)\sin(\omega x) dx  \nonumber\\
&& + \cos(\omega hk) \int_{0}^\infty f_n(x)\cos(\omega x) dx  + \sin(\omega hk) \int_{0}^\infty f_n(x)\sin(\omega x) dx  \nonumber\\
&=& \cos(\omega hk) \int_{-\infty}^\infty f_n(x)\cos(\omega x) dx + 2  \sin(\omega hk) \int_{0}^\infty f_n(x)\sin(\omega x) dx  \nonumber \\
&=& \cos(\omega hk) \real\left(\hat{F}(\omega)\right)  + 2  \sin(\omega hk) \int_{0}^\infty f_n(x)\sin(\omega x) dx  \label{eq:combination} \;.
\end{eqnarray}
Consider the Laplace transformation of $f_n(x)$,
\begin{eqnarray}
  \tilde{F}(s) = \mathcal{L}\{f_n(x)\} = \int_{0}^\infty f_n(x) \exp(-sx) dx \;.
\label{eq:LaplaceTransform}
\end{eqnarray}
The imaginary part of \eqref{eq:LaplaceTransform} is 
\begin{equation*}
 \imag (\tilde{F}(s)) = - \int_{0}^\infty f_n(x) \sin(sx) dx \;.
\end{equation*}
From this \eqref{eq:combination} can be expressed by
\begin{equation*}
\hat{F}(\omega) =  \cos(\omega hk) \real\left(\hat{F}(\omega)\right) - 2  \sin(\omega hk) \imag\left(\tilde{F}(\omega)\right) \;, 
\end{equation*}
where $\tilde{F}(\omega)$ is the Laplace transformation. 


\bibliographystyle{eptcs}

\bibliography{lit}

\end{document}